\documentclass[aps,pre,showpacs,floats,twocolumn,floats,superscriptaddress,floatfix]{revtex4-1}
\usepackage{bm}
\usepackage{url}
\usepackage{times}
\usepackage{verbatim}
\usepackage{theorem}
\usepackage{makeidx}
\usepackage{amsmath}
\usepackage{siunitx}
\usepackage{soul}
\usepackage{xy}
\usepackage{amssymb}
\usepackage{showlabels}
\usepackage{graphicx}
\usepackage{color}
\usepackage{epsfig}

\usepackage{subfigure}  
\usepackage{hyperref}   

\newcommand{\ictsaddress}{International Centre for
  Theoretical Sciences, Tata Institute of Fundamental Research,
  Bangalore 560089, India}
\newcommand{\iiseraddress}{Indian Institute for Science Education and Research, Pune, 411008, India }

\newcommand{\nordita}{Nordita, KTH Royal Institute of Technology and Stockholm University, 10691 Stockholm, Sweden}
\newcommand{\uniroma}{Department of Physics and INFN, University of Rome Tor Vergata,Via della Ricerca Scientifica 1, 00133 Rome, Italy}

\begin{document}
\title{Understanding Droplet Collisions through a Model Flow: Insights from a Burgers Vortex}
\author{Lokahith Agasthya}
\email{lnagasthya@gmail.com}
\affiliation{\iiseraddress}
\affiliation{\ictsaddress}
\affiliation{\uniroma}
\author{Jason R. Picardo}
\email{jrpicardo@icts.res.in}
\affiliation{\ictsaddress}
\author{S. Ravichandran}
\email{ravichandran@su.se}
\affiliation{\nordita}
\author{Rama Govindarajan}
\email{rama@icts.res.in.com}
\affiliation{\ictsaddress}
\author{Samriddhi Sankar Ray}
\email{samriddhisankarray@gmail.com}
\affiliation{\ictsaddress}
\begin{abstract}

	We investigate the role of intense vortical structures,
	{similar to those in a turbulent flow}, in enhancing collisions (and coalescences) which lead to the
	formation of large aggregates in particle-laden flows. {By
	using a Burgers vortex model, we} show, in particular, that vortex
	stretching significantly enhances sharp inhomogeneities in spatial
	particle densities, related to the rapid ejection of particles from
	intense vortices.  Furthermore our work shows how such spatial
	clustering leads to an enhancement of collision rates and extreme
	statistics of collisional velocities. We also study the role of
	poly-disperse suspensions in this enhancement. Our work uncovers an
	important principle which, {if valid for realistic turbulent
	flows, may be a factor in} how small nuclei water droplets in warm
	clouds can aggregate to sizes large enough to trigger rain.

\end{abstract}

\maketitle

\section{Introduction}

The transport of small, often spherical, particles in turbulent flows are
central to several processes in nature and industry.  These phenomena span
across several orders of length and time scales and are of importance for problems
in statistical physics, fluid dynamics, geophysics as well as astrophysics.
More often than not, such particles are neither massless nor point-like
(tracers): Hence our understanding of the anomalous nature of Lagrangian
turbulence is often inadequate when dealing with small, but finite-sized, heavy
particles~\cite{Bec2003}.  This is because such \textit{inertial}
particles---embryonic rain drops in a warm
cloud~\cite{doi:10.1175/1520-0450(1993)032<0608:WRIAOO>2.0.CO;2,PhysRevLett.112.184501,PhysRevE.93.031102},
pollen grains and pollutants~\cite{borgas_sawford_1994,Rama-review}, or even
planetesimals in a dusty circumstellar disk of gas~\cite{Planets-Nature}---behave very differently from tracers in a
flow. A key \textit{difference} between the dynamics of tracers and inertial
particles is \textit{dissipation}: The phase space for tracers is conserved
whereas that associated with inertial particles shrinks~\cite{Bec2003}.  A
dramatic visual manifestation of this is seen, e.g., in a snapshot of an
ensemble of inertial particles in a flow. Such configurations show an
inhomogeneous, preferential concentration of the particles in certain regions
of the flow unlike the case of tracers which distribute
homogeneously~\cite{doi:10.1063/1.858045,wang_maxey_1993,WOOD20051220,doi:10.1146/annurev.fluid.010908.165243}. 

Amongst the several problems of interest in this field, a recent focus has been
on the issues of
collisions~\cite{doi:10.1146/annurev.fluid.35.101101.161125,Bec2005,Falkovich2007,PhysRevE.87.063013,Pumir2016,JamesRay,doi:10.1063/1.4900848,Picardo2018-coll},
coalescences~\cite{PhysRevE.93.031102,Deepu2017,doi:10.1175/1520-0450(1993)032<0608:WRIAOO>2.0.CO;2,doi:10.1146/annurev.fluid.35.101101.161125}
and settling~\cite{PhysRevLett.112.184501,maxey_1987} of such particles in a
turbulent flow. Although such questions are important for fluid dynamics and
the statistical physics of non-equilibrium transport and aggregation problems,
there has also been a greater appreciation for its implication in
understanding, e.g., the microphysics of droplet growth leading to
precipitation in warm clouds. This is because the growth of water droplets from
small aerosols is, beyond a certain size, dominated by coalescence. Hence the
role of turbulent fluctuations in broadening the size distribution of droplets,
beginning with, e.g., a mono-disperse suspension of particles, deserves
attention.   

How effective is turbulence-driven coalescence in triggering the growth of
large objects? The answer to this question must depend on the way particles
approach each other. This rate of approach has its origins in two distinct
features of particle dynamics, namely, preferential concentration leading to an
inhomogeneous distribution of particles, and sling or caustic
effects~\cite{Falkovich2002,Wilkinson2006,Bewley2013} which cause heavy
particles, of suitable sizes, to collide with unusually large velocity
differences. A mechanical interpretation of both these effects can be traced
back to the proliferation of vortices in a three-dimensional flow: Vortices in
a flow act like centrifuges, expelling heavy particles from their core. Hence
particles are seen to cluster in regions of low rotation and high strain. Furthermore, expelled
particles which are sufficiently large (yet smaller than the Kolmogorov scale
of the flow) can meet other particles with arbitrary (and hence often large)
velocity differences, and thereby collide rapidly. However, the precise 
role of the curious inter-twined geometry of straining and vortical regions in particle-collisions has been examined 
in fully developed three-dimensional turbulent flows, via direct numerical simulations, 
only recently~\cite{Picardo2018-coll}. The probability of
coalescence upon collision of two droplets is a complex issue, beyond
the scope of this paper.

The question remains, however, of the precise importance of vortical structures
in facilitating droplet growth, especially when the carrier flows are strongly
turbulent.  A clear, definitive answer is difficult for two principal reasons.
Firstly, full-resolved numerical simulations can rarely reach Reynolds numbers
which are large enough for us to isolate the role of vortices with a high
degree of satisfaction. Secondly, inertial particles have non-trivial
correlations with the advecting flow: This makes it hard to isolate the effect
of vortical ejections of particles  on collisions and coalescence.  

Nevertheless, this question is an important one, especially if we want to
understand natural phenomena, such as droplet growth in warm clouds. Such
clouds are highly turbulent, and the role of vortices needs to be examined with
care. However, such Reynolds numbers are beyond the reach of fully-resolved
simulations, based on the Navier-Stokes equations. Hence the need to
examine reduced models to shed light on this phenomenon. 

{A first step in this direction, by Ravichandran and
Govindarajan~\cite{Ravichandran2015} (and later developed in
Ref.~\cite{Deepu2017}), was the use of particles interacting \textit{only} with
a two-dimensional stationary point or Gaussian vortices. Their work uncovered two
important aspects of vortex-particle interactions.  Firstly, it was shown that
caustics, defined as the co-existence at some physical location of two
particles with different velocities,  occur only if (a) at least one of the
particles started within a critical radius $r_{cr}$ of the centre of the vortex
and (b) the particle Stokes number, to be defined precisely later, is greater
than a critical value $St_{cr}$. Secondly, as a consequence of this, particles
starting near a vortex undergo significantly more collisions as they are expelled out, than particles
starting far from the vortex.}

In this paper, we extend these ideas concretely in a three-dimensional set-up with stretched Burgers vortices. In particular,
we elucidate the complementary roles of axial straining and intense vorticity in
enhancing collision rates, primarily through the mechanism of
\textit{slings}~\cite{Falkovich2007}. 

Before we get into a detailed description of the model and our results, it
would be useful at this stage to stress that the Burgers vortex, despite its
simplicity and limited connection with the phenomenology of three-dimensional
turbulence, has been an important testing ground for several ideas for the
mathematical and fluid dynamical aspects of the Navier-Stokes equation.  The
Burgers vortex~\cite{BURGERS1948171}, being an exact solution of the
Navier-Stokes equation, is an important model for self-similar flows which are
stationary, where the intensification of vorticity due to the axial straining flow---leading to sharp columnar
structures---is balanced by viscous diffusion. This is indeed reminiscent of
extended vorticity filaments seen in turbulent flows, both
numerically~\citep{she1990,Jimenez1998,ishihara2007} and in
experiments~\citep{Douady1991}. Consequently, it serves as an important model
for vortex stretching---a uniquely three-dimensional phenomena---and hence for
fundamental mathematical studies of (dissipative) Euler and Navier-Stokes
solutions~\cite{doi:10.1063/1.863957,GIBBON1999497,Galanti:1997,Gibbon:1997:10.1063/1.869186}. {Moreover, ensembles of such model vortices have been shown to provide a reasonable approximation, and hence a testing ground, for realistic turbulent flows~\cite{Chapman}. This has motivated earlier authors to use such models for understanding problems of Lagrangian turbulence~\cite{Wilczek} as well as turbulent transport of heavy inertial particles~\cite{Marcu1998,Hill}. In a similar spirit, we analyze the relative motion of droplets suspended in and around a Burgers vortex, in order to gain insight into the enhancement of droplet collisions by intense vortical structures in turbulent flows.
}
 
The rest of the paper is organised as follows. In Sec.~\ref{model}, we describe
our model of the Burgers vortex interacting with inertial particles, as well as
the parameters that we use.  We then describe the results we obtain from our
detailed numerical simulations in Sec.~\ref{results} and conclude with
discussions---especially the relevance of these results to the problem of
droplet growth in warm clouds---in Sec.~\ref{conclusions}.

\section{The Burgers Vortex Model}
\label{model}

We consider a single, cylindrically symmetric, three-dimensional,
\textit{stationary}, model vortex such that the vorticity is maximum at its
centre and falls off as we move away from the core. Such a Burgers
vortex~\cite{BURGERS1948171}, centered at the origin and aligned along the
${\bf {\hat z}}$-axis (without any loss of generality), and characterised by
its circulation $\Gamma$ and radius $r_{\rm core}$, is best described through
the velocity field written in cylindrical coordinates (and in component form): 
\begin{subequations}
\label{eq:Burgers}
\begin{align}
u_{\theta} &= \frac{\Gamma}{2 \pi r} \left(1 - \exp\left(-\frac{r^2}{r_{\rm core}^2}\right)\right);\\
u_r &=  -\sigma r; \\
u_z &= 2 \sigma z. 
\end{align}
\end{subequations}
This structure of the velocity field, along with the stretching coefficient
$\sigma$, ensures outward stretching along the ${\bf
{\hat z}}$-axis, accompanied by a radially inward component as demanded by incompressibility. Therefore, for small values of $r$, i.e., close to the vortex
core, the tangential velocity $u_{\theta}$ dominates and the flow is mostly
rotational; at large distances $r$, however, the radially-inward, axially-outward straining flow dominates.  Such a flow~\eqref{eq:Burgers} leads to a vorticity
field, which has a single non-trivial (Gaussian) $z$-component, given
by~\cite{Davidson}  
\begin{equation}
\omega_z = \frac{\Gamma}{\pi r_{\rm core}^2} \exp\left(-\frac{r^2}{r_{\rm core}^2}\right).
\end{equation}

We now seed heavy inertial particles near the origin, with an initial
velocity that matches the velocity of the fluid at that position.  The
dynamics of such small, dense, spherical particles is defined by the linear
Stokes drag model:
\begin{eqnarray} \frac{d{\bf
x}}{dt} &=& {\bf v}; \nonumber \\ \frac{d{\bf v}}{dt} &=&  - \frac{{\bf v} - {\bf u}}{\tau_p}.
\label{eq:simp_Max}
\end{eqnarray}
This set of equations, derived from the so-called Maxey-Riley
equation~\cite{Maxey1983} (see also Ref.~\cite{Rama-review}), allows
us to obtain the position ${\bf x}$ and velocity ${\bf v}$ of a given particle
at any instant of time.  The Stokes time $\tau_p$---a measure of the inertia of
the particle---sets the time-scale needed by the particle to relax to the
velocity of the carrier fluid. In most studies of turbulent transport, this measure of
inertia is best expressed through the non-dimensional Stokes number $St =
\tau_p/\tau_\eta$, where $\tau_\eta$ is the so-called Kolmogorov (small-scale)
time-scale of an advecting {\it turbulent} flow. Although we examine a model
flow, it is useful to adapt this definition to make meaningful comparisons to
warm clouds, where such a vortex-particle system mimics their dominant vortical
structures---because of extreme Reynolds numbers---interacting with nuclei
droplets. In our numerical simulations, we use a second-order Runge-Kutta
scheme, with a time step $\delta t = 10^{-3}$, and a few values of $\tau_p$ (see
Table~\ref{tab:para}), to solve the equations of motion for the particles. For
well-converged statistics, we use $N_p = 200,000$ particles for each of our
simulations. It is important to point out that we report results for a rather
narrow range of Stokes numbers (0.03 to 0.1). This is because of two
reasons. Firstly, the rather subtle effect that we are interested in is really
dominant for this range of $St$; the effect is inconsequential for tracers, while
for finitely large values of $St$ the ballistic motion of particles
overwhelms the role played by the geometry of our model flow. Secondly, these
Stokes numbers are realistic for droplet-nuclei in settings such as warm
clouds, whose physics is part of the motivation for this study.

\begin{table}
    \begin{center}
    \begin{tabular}{|c|c|c|c|}
    \hline
    \multicolumn{3}{|c|}{Fluid (Burgers Vortex)}& Particles \\ 
    \hline
    Type & $r_{\rm core}$ & $\sigma$ & $St$ \\
    \hline
    & $0.2$ & $0$ & $0.03$\\
    \cline{2-4}
    & 0.2 & 0.08 & 0.03 \\
    \cline{2-4}
    & $0.2$ & $0.08$ & $0.06$ \\
    \cline{2-4}
    $V_1$& $0.2$ & $0.08$ & $0.1$  \\
    \cline{2-4}    
    & $0.2$ & $0.25$ & $0.03$ \\
    \cline{2-4}
    & 0.2 & 0.3 & 0.03 \\
    \cline{2-4}
    & $0.2$ & $1.0$ & $0.03$ \\
    \hline
    & $0.4$ & $0$ & $0.03$\\
    \cline{2-4}
    & 0.4 & 0.02 & 0.03\\
    \cline{2-4}
    $V_2$ & $0.4$ & $0.08$ & $0.03$\\
    \cline{2-4}
    & $0.4$ & $0.25$ & $0.03$\\
    \cline{2-4}
    & $0.4$ & $1.0 $ & $0.03$\\
    \hline
    
    \end{tabular}
    \end{center}
        \caption{Representative list of parameters that have been used in our numerical simulations and reported in this paper.} 
     \label{tab:para}
\end{table}

\begin{figure*}
\includegraphics[width=0.49\textwidth, keepaspectratio]{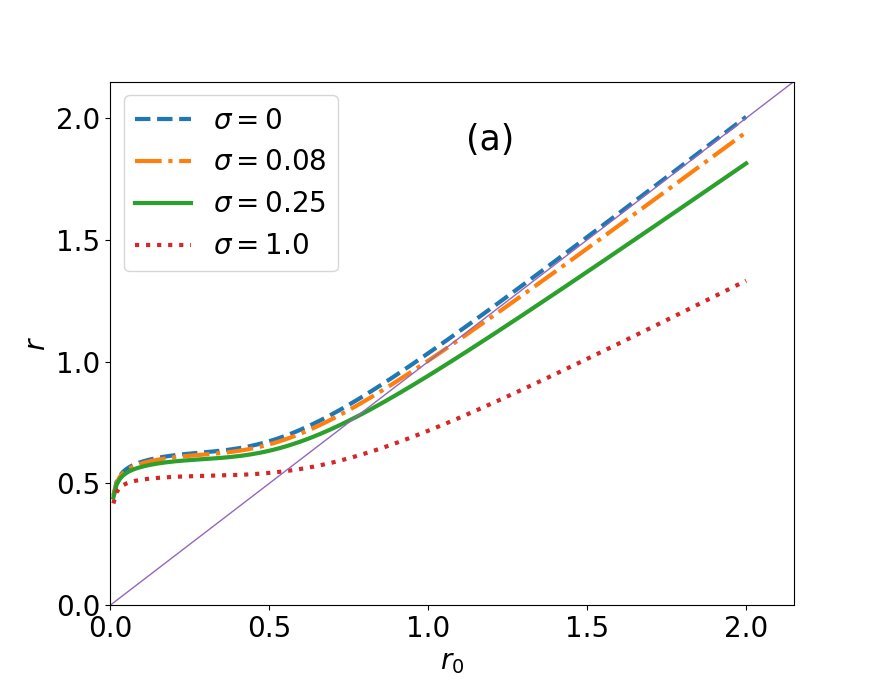}
\includegraphics[width=0.49\textwidth, keepaspectratio]{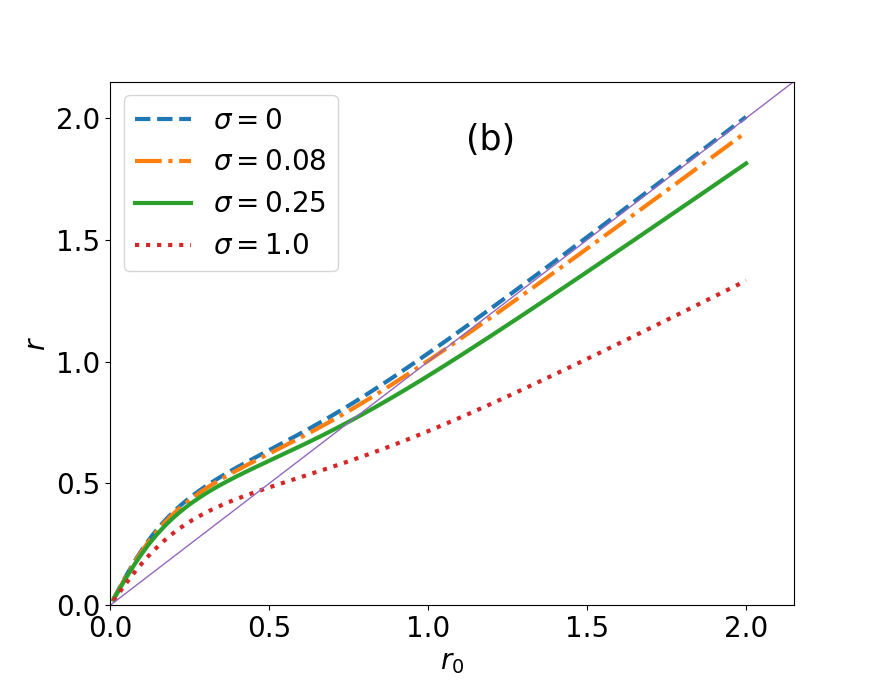}
\caption{(color online) Representative plots of the final radial distance $r$ as a
function of the initial radial distance $r_0$ for particles with $St = 0.03$ at
a short time $t = 0.6\tau_\eta$ for a (a) strong ($r_{\rm core} = 0.2$) and (b)
weak ($r_{\rm core} = 0.4$) vortex. We show data for different flow geometries
characterised by $\sigma$, and the unbroken (purple) diagonal line is a guide to
the eye to differentiate particles which are centrifuged out ($r>r_0$)
from those which are drawn in ($r<r_0$).}
\label{fig:radial}
\end{figure*}

Measurements suggest a spread of the mean-energy-dissipation rates as well as the
Reynolds numbers in a turbulent warm cloud.  We choose a typical set of values
for the  mean-energy-dissipation rate $\epsilon = 0.01
\SI{}{\meter\squared\per\second\cubed}$, and Taylor Reynolds number $Re_{\lambda} = 5000$, as
well as the kinematic viscosity $\nu = \SI{1.48e-5}{\meter\squared\per\second}$
for air. It is important to keep in mind that, given the nature of our work, the
precise combination of these numbers does not fundamentally alter our results and
the conclusions that we draw from them. Having measurements of
$\nu$ and $\epsilon$, it is trivial now to obtain the Kolmogorov length-scale $\eta$
and time-scale $\tau_\eta$, as well as  the spatially-averaged {\it enstrophy}
$\langle\omega^2\rangle = \frac{\epsilon}{\nu}$, associated with the turbulent
flow in such clouds. 

These are, of course, intensely turbulent flows and our understanding of such
systems suggests proliferation of intense, sharp columnar vortical
structures~\cite{Jimenez1998}, which empirically are regions of vorticity
greater than $\sqrt{\langle \omega^2 \rangle Re_{\lambda}}$.  In this paper, it is this
peculiar aspect of highly turbulent regimes that we replicate through the
Burgers vortex defined earlier. We choose 2 different sets of Burgers
vortices---namely a {\it small core} and {\it large core} vortex, corresponding to two
different values of $r_{\rm core}$ (and the associated circulation $\Gamma$ and
stretching coefficient $\sigma$). {Our choice of parameters for our simulations
is listed in Table 1, and how these dimensionless parameters are obtained
from the various dimensional measurements in. e.g., a turbulent cloud, is described in
Appendix A.} 

Finally, we seed particles with several different Stokes numbers, including the
typical value of $St = 0.03$ as commonly seen in nuclei aerosols. 

\section{Results}
\label{results}

To understand the dynamics of inertial particles near a strong vortex, it is
best to begin by examining single particle trajectories, starting out at an initial
radial distance $r_0$ (measured from the axis of the vortex). The time
evolution of such a particle, in particular its radial distance $r(t)$,
ought to depend on both $r_0$ and the stretching coefficient $\sigma$, which
counters the centrifugal effect of the vortex on the particles. In
Fig.~\ref{fig:radial}, we show representative plots of the radial distance $r$
for a particle with $St = 0.03$ at time $t = 0.6\tau_\eta$,  for both the small
($r_{\rm core} = 0.2$) and large ($r_{\rm core} = 0.4$) core vortices (panels
(a) and (b), respectively).  Our results clearly show that although we expect
the particles to be centrifuged out of the vortex, for sufficiently large
$\sigma$, and $r_0$, the particles actually come closer to the vortex axis. To
make this clear, we show, in Fig.~\ref{fig:radial}, a diagonal line 
which serves as a guide to the eye: Portions of the curves lying above this
diagonal represent particles which, for a given $r_0$ and $\sigma$, centrifuge
out and move away from the vortex-axis; the part of the curve lying below
this diagonal correspond to the cases where $r_0$ and $\sigma$ conspire to pull
the particles into the vortex. Particles with a sufficiently small
$r_0$ always end up moving away from the vortex and their radial distance at short
times seem to be independent of $\sigma$, but influenced by the strength of the
vortex: a smaller $r_{\rm core}$ (or a more \textit{intense} vortex) leads to a
more violent ejection. Interestingly, though, for a sufficiently large value of
$r_0$, the role of $\sigma$ becomes non-trivial. Hence, for any given initial
separation $r_0$, particles may well be brought \textit{inwards} if the
stretching of the flow $\sigma$ is strong enough. Of course, at very long times
all particles, regardless of $r_0$, will asymptote to a distance $r_\star$ determined by
$\Gamma$ and $\sigma$, where outward centrifugal forces are balanced by inward drag 
from the stretching flow~\citep{Marcu1998}. We have checked that this effect is
indeed independent of the Stokes numbers that we consider in this paper. 

It is worth making a final comment on Fig.~\ref{fig:radial}. For strong
vortices (panel (a)), particles which start at different, but reasonably short
distances $r_0 \lessapprox r_{\rm core}$ end up roughly at the same $r =
r_\star \gtrsim r_{\rm core}$, as seen by the curves plateauing out.  This in
turn implies that particles which start closer to the center of the vortex have
a higher speed of ejection than those that start further away. This effect 
naturally weakens when the strength of the vortex diminishes, as is evidenced by the absence of a similar plateau in Fig.~\ref{fig:radial}(b).
{The differential ejection of particles, in case of a strong vortex, enables particles which begin near the vortex to overtake, in finite time, those that begin further away, resulting in the formation of caustics~\cite{Ravichandran2015}.}

\begin{figure}
\includegraphics[width=0.49\textwidth, keepaspectratio]{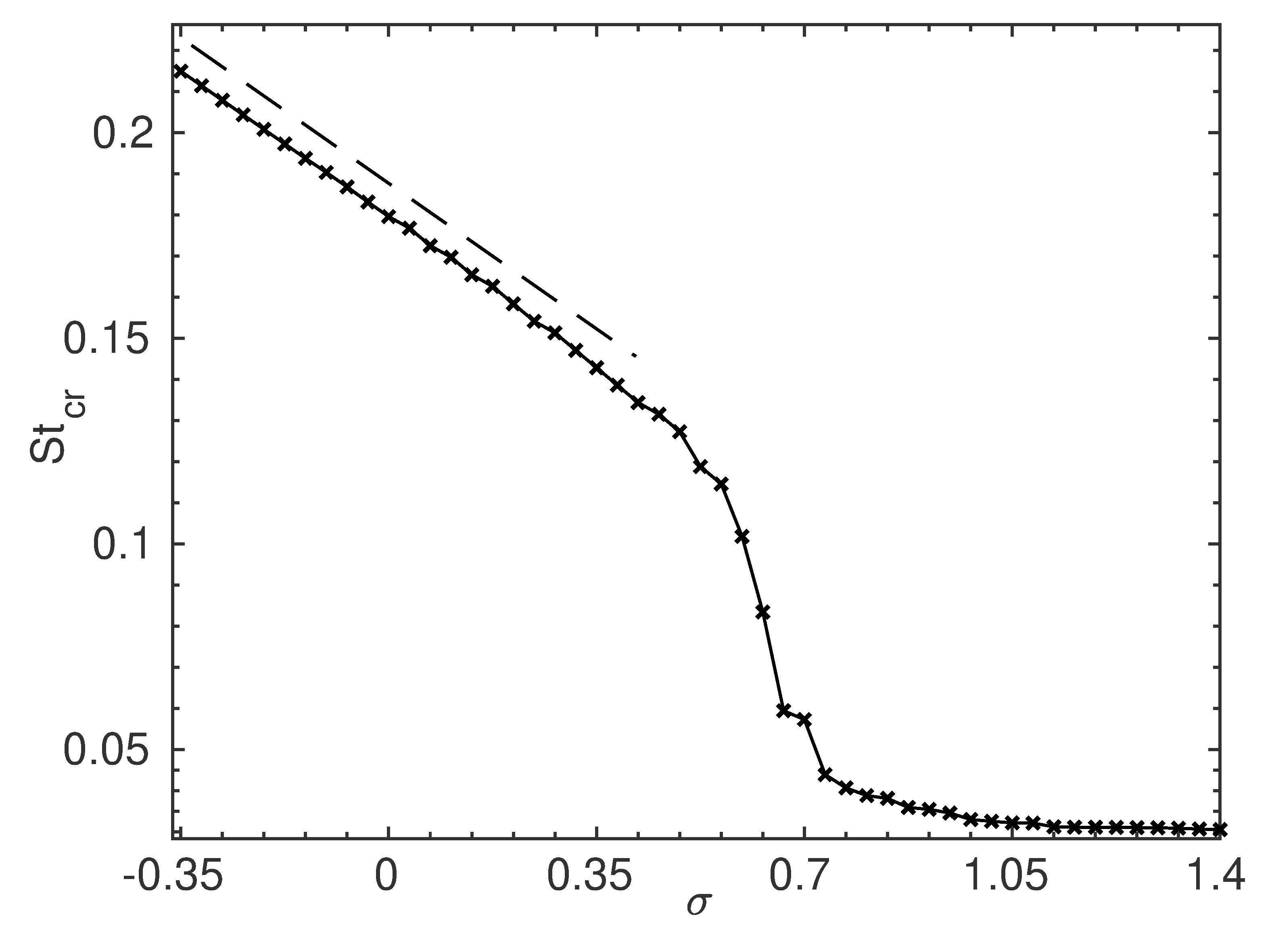}
\caption{The critical Stokes (see text for definition) as a function of the nondimensional stretching rate.
For small stretching rates, the critical Stokes falls linearly with increasing rate (the slope of the dashed line is $-1/10$). There also seems to be
an asymptotic value at high stretching rates.}
\label{fig:Stcr}
\end{figure}

{To understand how vortex stretching impacts the formation of caustics, we calculate the critical Stokes number,
$St_{cr} = \tau_{p,cr} / \tau_\eta$ (for the mono-disperse case). Given the vortex strength and the stretching rate, particles with $St < St_{cr}$ will not undergo caustics. The variation of $St_{cr}$ with the stretching rate thus quantifies the effect of stretching on caustic-formation, and is shown for the present value of $Re_\lambda$ in Fig.~\ref{fig:Stcr}.}  A scale-free version
(again following the scaling used in Ref.~\citep{Ravichandran2015}) where the
dependence on the Taylor Reynolds number is apparent is shown in the Appendix.  
{The decrease in the critical Stokes number with increasing 
stretching rate suggests that, for a given particle size,  a vortex that is stretched will produce more 
caustics than a vortex that is not.}
We
shall see below that, apart from this, the collisions at a given Stokes number
are enhanced by stretching as well.

All of this suggests that the number density of particles $\Phi$, at some time
$t$, should depend sensitively on $r$. We define this density as the ratio of
the number of particles $N(r)$ at a given $r$ (suitably normalised and binned)
to the total number of particles $N_p$, {\it i.e.}, $$ \Phi =
\frac{1}{2 \pi r h} \frac{N(r)}{N_p}$$ where $h$ is the height of the cylinder in which 
the particles are uniformly distributed initially. 

\begin{figure}
\includegraphics[width=1.0\columnwidth, keepaspectratio]{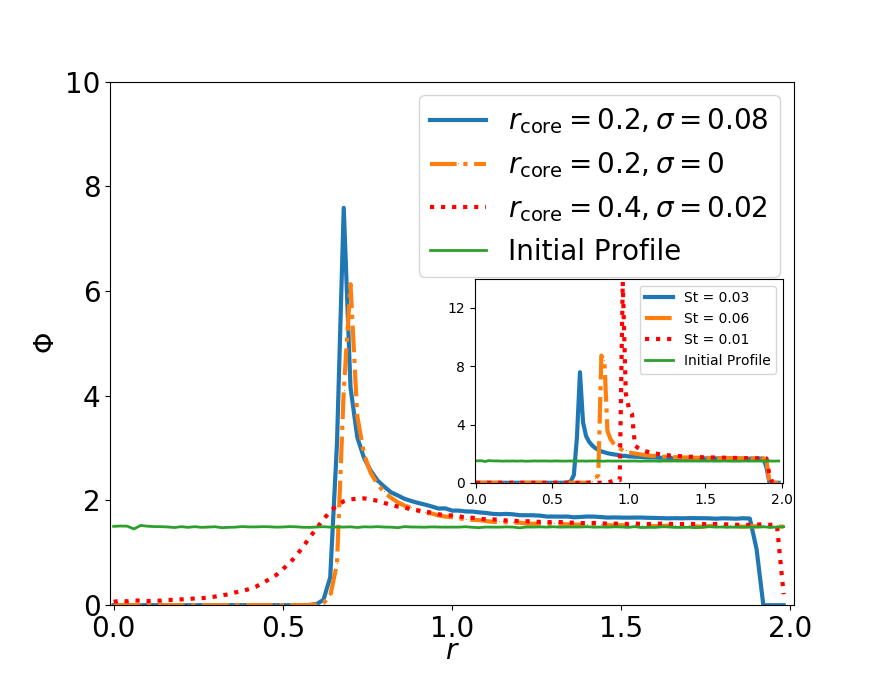}
	\caption{(color online) Representative plot of the particle number density $\Phi$, for $St = 0.03$, 
	as a function of $r$ at time $t
	= 0.6\tau_\eta$ for different flow geometries (see legend). 
	The effect of particle inertia is illustrated in the inset which 
        shows $\Phi$ \textit{vs} $r$, for a given flow geometry ($r_{\rm core} = 0.2$, $\sigma = 0.08$), 
        for particles with $St = 0.03$, $St = 0.06$
	and $St = 0.1$. The horizontal, green unbroken line is the
	initial uniform density profile.}
\label{fig:Phi}
\end{figure}

{The number density of particles $\Phi$ is a non-stationary measure; 
however since we want to capture the collision statistics dominated by vortical expulsion, 
it is enough to measure this quantity at times that are just long enough for most of the particles to have left the vortical core and accumulate near $r_*$. This is because it is in this setting 
that collisions will be dominated by the vortical expulsion of
particles. Hence, in} Fig.~\ref{fig:Phi}, we show $\Phi$ (for $St = 0.03$ and
averaged over {100} different initial realisations of the particles) as a function
of $r$, at $t = 0.6\tau_\eta$, for different vortices and stretching rates. 
{The time chosen here satisfies the constraint discussed earlier, without being so large as to allow particles that were initially far from the vortex to become concentrated at $r_*$, because of the inward flow due to straining. 
(We have checked that our conclusions
and results (qualitatively) are unchanged for slight variations to the
waiting time.)}

We see clearly that starting from an initial profile in which particles are
uniformly distributed in space, the particles evacuate rapidly from regions
close to the core of the vortex. (The case of $\sigma =0$ is validated against
Ravichandran and Govindarajan~\citep{Ravichandran2015}.) Therefore, at some
later time ($t = 0.6\tau_\eta$), $\Phi$ is essentially 0 at small values of $r$
before sharply peaking at $r_\star$.  The onset of this peak corresponds to the
flat region already noticed in Fig.~\ref{fig:radial}.  Understandably, $\Phi$
falls off---less rapidly---as $r > r_\star$. At later times, the density
profile is qualitatively the same, but with a sharper peak that is shifted
outward. Eventually, the profile asymptotes to a delta function, as $t\to
\infty$, at a value of $r_\star$ where the inward flow-drag, due to stretching,
is balanced by the centrifugal action of the vortex~\cite{Marcu1998}.

In Fig.~\ref{fig:Phi} we also notice two distinct effects. The peak in the
density profile, at any given time, is significantly sharper and more pronounced in the presence of
stretching. This is because stretching counters the expulsion process due to the
vortex and hence constraints particles to concentrate more sharply around
$r_\star$. Furthermore, for the weaker vortex, not only is the density profile
less sharp, it also shows clearly that evacuation of particles
from the vortex core is incomplete, as is also suggested by the lack of a clear plateau in
Fig.~\ref{fig:radial}(b). 

What role does the Stokes number of the particles play in all of this? In the
inset of Fig.\ref{fig:Phi} we answer this question precisely, by showing the
density profile for different values of $St$, for a given flow geometry. As we
would expect, the critical distance $r_\star$ is a monotonically increasing
function of the Stokes number, even in the limit $t \to \infty$. 

This distinct and sharp profile of $\Phi$ and the differential radial
evacuation of particles from the core of the vortex must have an important
bearing on the statistics of collisions---and hence coalescences---of inertial
particles.  In particular, it seems reasonable to assume that (a) the sharply
localised high value of $\Phi$---and the corresponding reduction in the average
inter-particle distance---ought to lead to a higher rate of collision; and
(b) the differential radial velocities of particles, which depend on their
initial distance from the vortex core as discussed above, would lead to
larger collisional velocities than in a more quiescent, vortex-free, region
of a flow. It therefore behooves us to test this conjecture in our model
system. We should keep in mind that our results are meaningful only for short
times: In an actual spatio-temporally varying turbulent flow, vortices are
spatially localised for very short times unlike our stationary model; hence we
are careful to consider only those collisions which occur at times shorter
than the Kolmogorov time-scale, in order for the insights we develop to have implications for realistic particle-laden flows. 

We use a standard algorithm for detecting particle
collisions~\cite{JamesRay,sundaram_collins_1997}, and reduce the
$N_p^2$ computational cost by dividing, at every time step, the spatial domain
into equal-sized grids.  By optimally choosing our grid sizes, we manage to
ensure that within one time step the particles do not cross more than one grid,
which guarantees that no collisions are missed even when the search is limited to the particle-containing grid and its nearest neighbours. After detecting a collision, the two particles are merged, conserving mass and momentum, into a new larger particle, located at the centre of mass of the colliding pair.

We initialise {200,000} particles, with radii $a =
\SI{2e-3}{} \approx \frac{\eta}{20}$, randomly in a small cubic domain centered at the
vortex. {This results in an initial particle volume fraction of $O(10^{-4})$ that is commensurate with the dilute situation encountered, e.g., in clouds.} The particles are then evolved according to \eqref{eq:simp_Max} up to a time $t =0.6\tau_\eta$. We simultaneously detect the number of collisions $\Theta$ which occur
per unit volume, as a function of the radial distance, by considering the
collisions which happen in concentric cylindrical shells parametrised by their radii
$r$. We, as before, averaged our data over {100} independent ensembles of initial
particle configurations.

\begin{figure}
\includegraphics[width=\columnwidth, keepaspectratio]{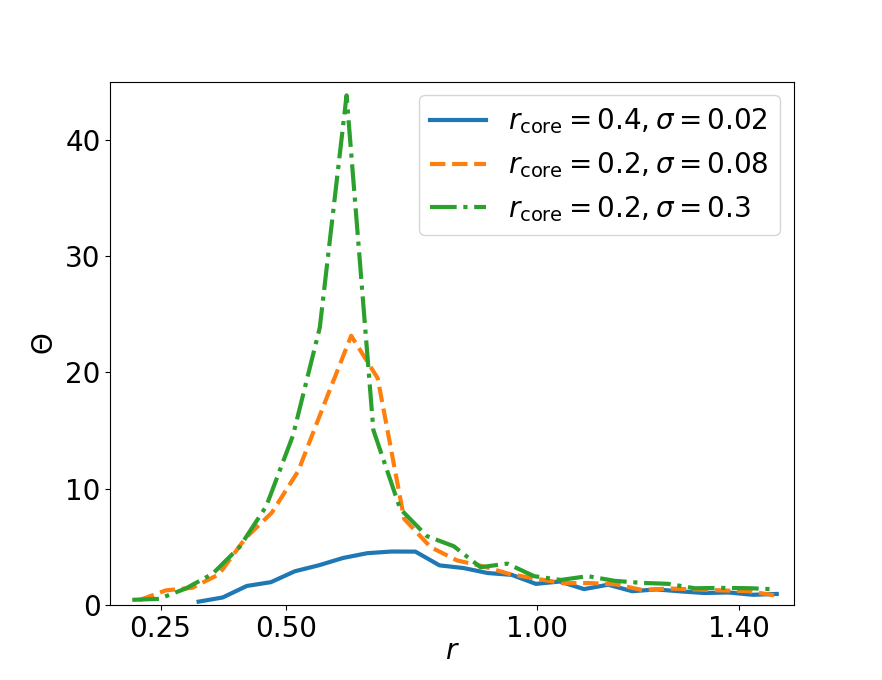}
\caption{(color online) {Plots} of the collision density $\Theta$ {(collisions occurring 
per unit volume)} as a function	of the radial distance $r$ for different flow geometries.}
\label{fig:theta}
\end{figure}

Figure~\ref{fig:theta} shows a plot of $\Theta$, as a function of $r$ for three
representative flow geometries, namely, ($r_{\rm core} = 0.4$, $\sigma =
0.02$), ($r_{\rm core} = 0.2$, $\sigma = 0.08$), and ($r_{\rm core} = 0.2$,
$\sigma = 0.3$). {(Our results from other sets of simulations with different parameters 
show the same qualitative behaviour.)}  We immediately notice more pronounced peaks in $\Theta$ for
the more intense vortices reminiscent of, and for the same reasons as, the effects we
had already seen in our measurement of the particle number density $\Phi$. Unlike the case of number
density, however, the spike in the measurement of $\Theta$, though clearly
concentrated around $r_\star$, is less sharp. 

We also see, in Fig.~\ref{fig:theta}, that an increase in the degree of
stretching $\sigma$---which accentuates the clustering around $r_\star$---leads to a considerable increase in the number of collisions.
However, could there be another explanation for this effect different from the
one we suggest---namely the outward centrifugal forces and the inward stretching flow
conspiring together?  Indeed, for a three-dimensional system such
as ours, it is plausible that the increase in $\Theta$ with $\sigma$ could be
simply due to the motion of particles along the $\hat{z}$-axis because of the
fluid velocity component $u_z = 2 \sigma z$. To test this hypothesis, we
perform simulations where we retained the radially inward flow unchanged but artificially set
$u_z = 0$. Surprisingly, we found that a suppression of the axial flow actually results in an overall increase of approximately 12\% in the peak value of $\Theta$, indicating that the $\hat{z}$-directed flow actually serves to inhibit collisions. In retrospect, this is understandable because the stretching flow pulls particles apart along the $\hat{z}$-direction, thereby reducing the local density and likelihood of collisions.

This is a rather interesting and important observation, which suggests that 
the majority of collisions, and
hence coalescences, are spatially localised because of two opposite, and yet
complementary effects: the centrifugal or \textit{sling} effects of a vortex
which evacuates particles \textit{away} from the vortex, and an inward flow
which actually brings-in particles \textit{towards} the
vortex. {Interesting as this observation may be, it is important 
to be cautious about how generic this effect could be in turbulent 
flows. Indeed this needs to be checked and studied in more realistic flows, involving direct
numerical simulations of the Navier-Stokes equations as well as gravitational
effects, to be able to conjecture that the rapid growth of aggregates in very
high Reynolds number flows, such as rain drops in warm clouds, owes its origin
to precisely this mechanism.} Our model system, nevertheless, allows us to
incorporate one additional feature which is especially relevant in the
understanding of droplet-growth through coalescences in clouds, namely,
poly-dispersity.

\begin{figure}
    \includegraphics[width=\columnwidth, keepaspectratio]{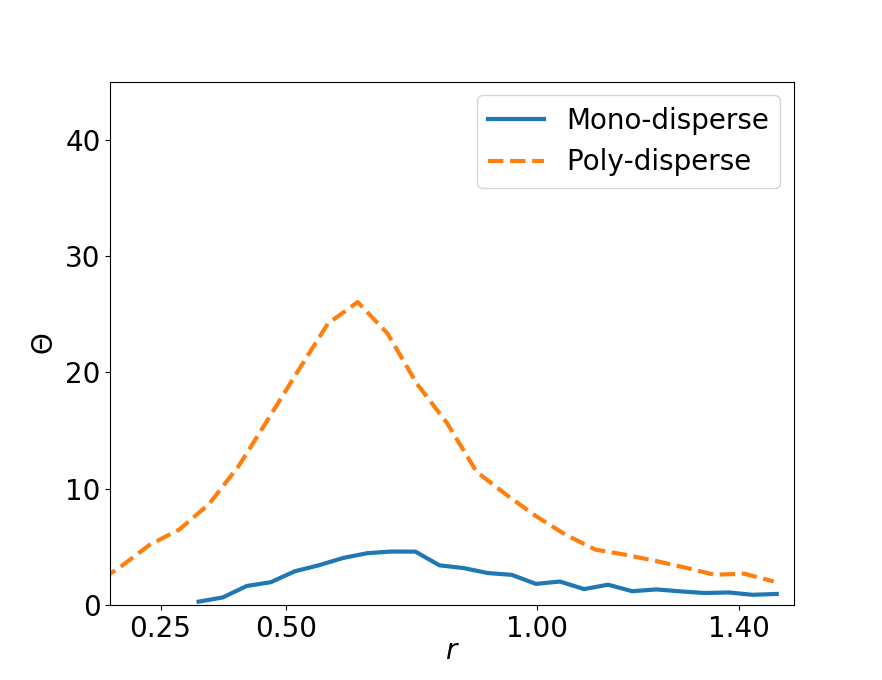}
	\caption{(color online) Representative plots of the collision number density $\Theta$, as a function of $r$,  
	illustrating the efficiency of a poly-disperse suspension over a mono-disperse one in facilitating 
	collisions, and hence coalescences. We show (see text) that poly-dispersity can overcome the effect of 
	a weak vortex and increase the value of $\Theta$.}
	\label{fig:poly}
\end{figure}

We now perform simulations with an ensemble of particles whose radii $a$ are
not constant (mono-disperse) but are actually taken from a Gaussian
distribution with a mean (radius) $a_0 = \SI{2e-3}{}$ and a standard deviation
which is $10\%$ of the mean. The Stokes numbers of these particles, naturally,
are also distributed, as $St \propto a^2$. 
This ensures that different particles within the same ensemble have different dynamics, because of their
differences in radii and therefore $St$.  The result of our investigations{, which extend 
recent work on the relative velocities of colliding particles in a bi-disperse 
suspension~\cite{JamesRay,Bhatnagar},} is
instructive. In Fig.~\ref{fig:poly} we show a representative plot where we
compare the collision rate $\Theta$ of a poly-disperse suspension  with that
obtained for a mono-disperse suspension (where particles have the same radius
as the mean radius in the poly-disperse case), in a flow with $r_{\rm core} =
0.4$ and $\sigma = 0.02$. Poly-dispersity dramatically
increases the number of collisions, in spite of the vortex in this case being
weak. This results from the distribution in ejection rates (each particle size
has a different clustering distance $r_*$), and shows a much higher value of
$\Theta$ than the mono-disperse case. An earlier study~\citep{Deepu2017} had found a dramatic dependence of the caustics radius on
small polydispersity. {These results now show how} this translates into
increased rates of collision. In fact, the $\Theta$ with polydisperse
particles is even larger than what is obtained in Fig.~\ref{fig:theta}, for
the mono-disperse case with a stronger vortex and a stronger inward flow
($r_{\rm core} = 0.2, \sigma = 0.08$).
{This result suggests that, at least in a model system, a spread in the size of droplets in
a suspension---as is the case in natural systems such as clouds---triggers a
more rapid growth in particle sizes through coalescences.} 

\begin{figure}
    \includegraphics[width=\columnwidth, keepaspectratio]{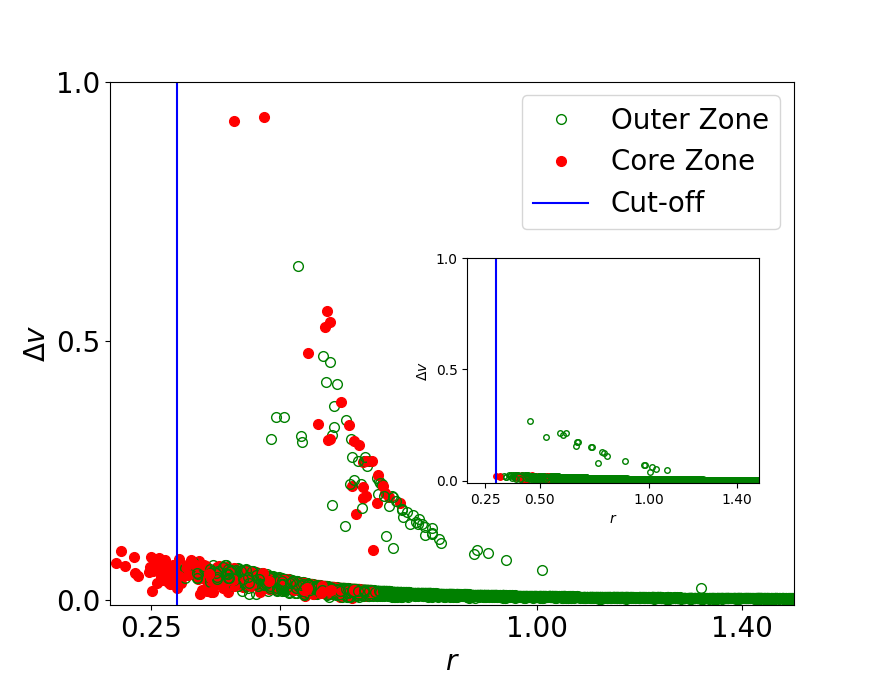}
	\caption{(color online) A scatter plot of the relative (collisional) velocities 
	for a small core and large core (inset) vortex corresponding to ($r_{\rm core} = 0.2, \sigma = 0.08$) and
	($r_{\rm core} = 0.4, \sigma = 0.02$) respectively. 
	The red, filled circles denote collisions where at least one 
	of the colliding particles have emerged from the heart (cut-off zone) of the vortex, denoted by the blue 
	vertical line (see text), whereas the green open circles are events where both  the particles 
	were initially outside this cut-off zone.}
	\label{fig:deltav}
\end{figure}

Our results so far, admittedly in a model flow, suggest a reasonably self-consistent picture 
of \textit{where} and \textit{how frequently} collisions occur in particle-laden flows dominated by columnar vortical structures. Before we conclude, it is useful to see what all of this 
means for the relative velocity $\Delta v$ of colliding droplets as well as the contribution of the Lagrangian 
history of particles in determining $\Delta v$. This is important not only for 
having an estimate of how \textit{fast} droplets hit each other, and hence whether these collisions lead to 
coalescences or fragmentation, but also crucial for modelling collision kernels 
in a mean field sense.

In Fig.~\ref{fig:deltav} we show a scatter plot of $\Delta v$ vs $r$
{corresponding to the collisions depicted in Fig.\ref{fig:theta}},
for both,the small core and the large core (inset) vortices.
We plot the value of $\Delta v$
with either a red (filled) circle or a green (open) circle; the red filled
circles are used when the trajectories of one or both the colliding particles start from a region very close to the vortex
core,
called the \textit{cut-off} zone for convenience and shown as a vertical blue line in the figure; the green open circles are meant for those collisions that
involve particles which were initially at positions outside
this cut-off zone. We choose the cut-off zone to be at $r = 0.3$, since, as we have seen
before (Fig.~\ref{fig:radial}a), this corresponds to the position where the centrifugal ejection is the
strongest for both the weak and strong vortices. 

Our results (Fig.~\ref{fig:deltav}) indicate an essentially bimodal distribution of 
$\Delta v$: There are rare collisions with relatively large collisional 
velocities, whereas most of the collisions have much milder impacts; extremal values of $\Delta v$ seem to be much larger for the stronger vortex than for the weaker one. Furthermore, especially for the stronger vortex, several high values
of $\Delta v$ correspond to collisions which occur due to one or both of the particles being 
violently ejected from the cut-off zone, and hence are marked in red filled circles. This disproportionate contribution to high-impact collisions from particles originating near the vortex core becomes clearer when we look at the numbers. 

Since our particles were initially distributed randomly, only 4\% of them had
initial positions $r_0$ within the vortex core. For the stronger
vortex these particles contribute to approximately 17\% of the collisions
whereas for the weaker flow this contribution (for the same initial
distribution) comes down to 2\%. This is visually illustrated by comparing the
number of red (filled) circles in Fig.~\ref{fig:deltav} and its inset. Indeed,
on average we have found that the relative velocity of collisions with
particles originating from within the cut-off zone are far higher than those involving particles 
outside it. This {strongly suggests} that rapid centrifugal ejection of particles, initially close to the centres of stretched tubular vortices,
lies at the heart of high velocity collisions in particle-laden, intensely turbulent flows.

\section{Conclusion}
\label{conclusions}

Our model, based on the premise that the singular structure of an intense vortex is a
key component of high Reynolds number flows, shows the extraordinary effect of
the interplay between the mechanics of centrifugal ejection and vortex stretching.
This shows up primarily in the spatial dependence of the particle number
density $\Phi$. This preferential clustering---which is enhanced for strong,
but stretched, vortices---has a direct bearing on collision frequencies
$\Theta$ and collisional velocities $\Delta v$.  We also examined the role of
poly-dispersity in our ensemble of particles and showed how, in fact, 
even a relatively narrow distribution of radii and Stokes times dramatically accelerates the rate of
collisions, and presumably therefore, of coalescences when the colliding objects are droplets. All of this suggests the importance of the
Lagrangian history of particle trajectories which lead us to uncover how
extreme events---yielding high values of $\Delta v$---can actually be traced
back to the proximity of particles to a vortex. 

Before we conclude, we must confront the question of the relevance of our
study---without overstating our case, keeping in mind that we study a model system---and its implications for realistic turbulent particle-laden flows.  To
answer this question, we recall that although our results are interesting and
generic, this work was in part motivated by the \textit{bottleneck problem}:
The question of how, starting from very small nuclei droplets that result from
condensation, large aggregates of rain-forming drops grow in a warm cloud. A
key ingredient in this process is coalescence, triggered by collisions, and the
role of turbulent mixing. {Given the very high Reynolds numbers encountered in such flows, the vorticity field is expected to be highly intermittent, with the most intense regions taking the form of stretched vortex tubes. It is therefore natural to ask what the precise role of intense vortex filaments is in the transport and collision-driven growth of particles. Hence our model.}

With abundant caution, dictated by the artificial and static nature of our flow, we 
investigated this question with detailed numerical simulations. Our results hint strongly 
at the possibility that the action of stretched vortices, as well as the narrow variation in 
size distribution of the nuclei droplets, enhance coalescence events to a great degree. 
It is within this context that this work assumes special importance. 

We hope that our results for this model system will trigger detailed direct numerical 
simulations to validate this mechanism in future studies. 

\section*{Acknowledgments}
SSR acknowledges the support of the Indo--French Center for Applied Mathematics (IFCAM) 
and the support of the DST (India) project ECR/2015/000361. The simulations were performed 
on the cluster {\it Mowgli}.
LA acknowledges the support of the INSPIRE Programme of the Department 
of Science and Technology, Government of India and the hospitality of ICTS-TIFR where this work was done. SR was employed at JNCASR, Bangalore, when the majority of this work was done.

\appendix
\setcounter{equation}{0}
\section{Choice of Parameters in our Numerical Simulations}
\label{a:num}

{A natural way in which the non-dimensional parameters for our simulations
could be chosen is to set the various scales in the problem 
through the intrinsic time-scale $\tau_{\eta} = r_{\rm core}^2 \Gamma^{-1}$ 
and Reynolds number $Re = 1/\sigma$ of the Burgers vortex model. However in this 
work we opt for a different strategy to make our results more testable against 
direct numerical simulations of highly turbulent flows or actual measurements 
in a physical system. We choose our model system to represent an intense vortex in a \textit{typical} warm 
cloud. Hence, we consider typical, measured values of the mean-energy-dissipation rate
$\epsilon = \SI{0.01}{\meter \squared \per \second \cubed }$, Taylor Reynolds number
$Re_{\lambda} = 5000$ and kinematic viscosity $\nu = \SI{1.48e-5}{\meter \squared \per
\second }$ in a turbulent warm cloud. This yields a Kolmogorov length scale
$\eta = \SI{7.54e-4}{\meter}$ and a Kolmogorov time scale $\tau_{\eta} =
\SI{3.8e-2}{\second}$.  Empirically, intense, sharp, columnar vortical
structures---which are mimicked by our Burgers vortex---are regions of vorticity
greater than $\sqrt{\langle \omega^2 \rangle Re_{\lambda}}$, where $\langle \omega^2
\rangle$ is the spatially averaged enstrophy given by $\langle \omega^2 \rangle
= \frac{\epsilon}{\nu}$~\citep{Jimenez1998}. An estimate of the width $l$ of such vortices 
suggests that $l \approx 10\eta$. By using standard (dimensional) relations between the vorticity, 
length scale and the velocity scale, $u =  l \sqrt{\langle \omega^2 \rangle Re_{\lambda}}$, we obtain the typical 
magnitude of the circulation of an intense vortex as $\Gamma = l^2 \sqrt{\langle \omega^2 \rangle Re_{\lambda}}$.}

\begin{figure}
\includegraphics[width=0.42\textwidth, keepaspectratio]{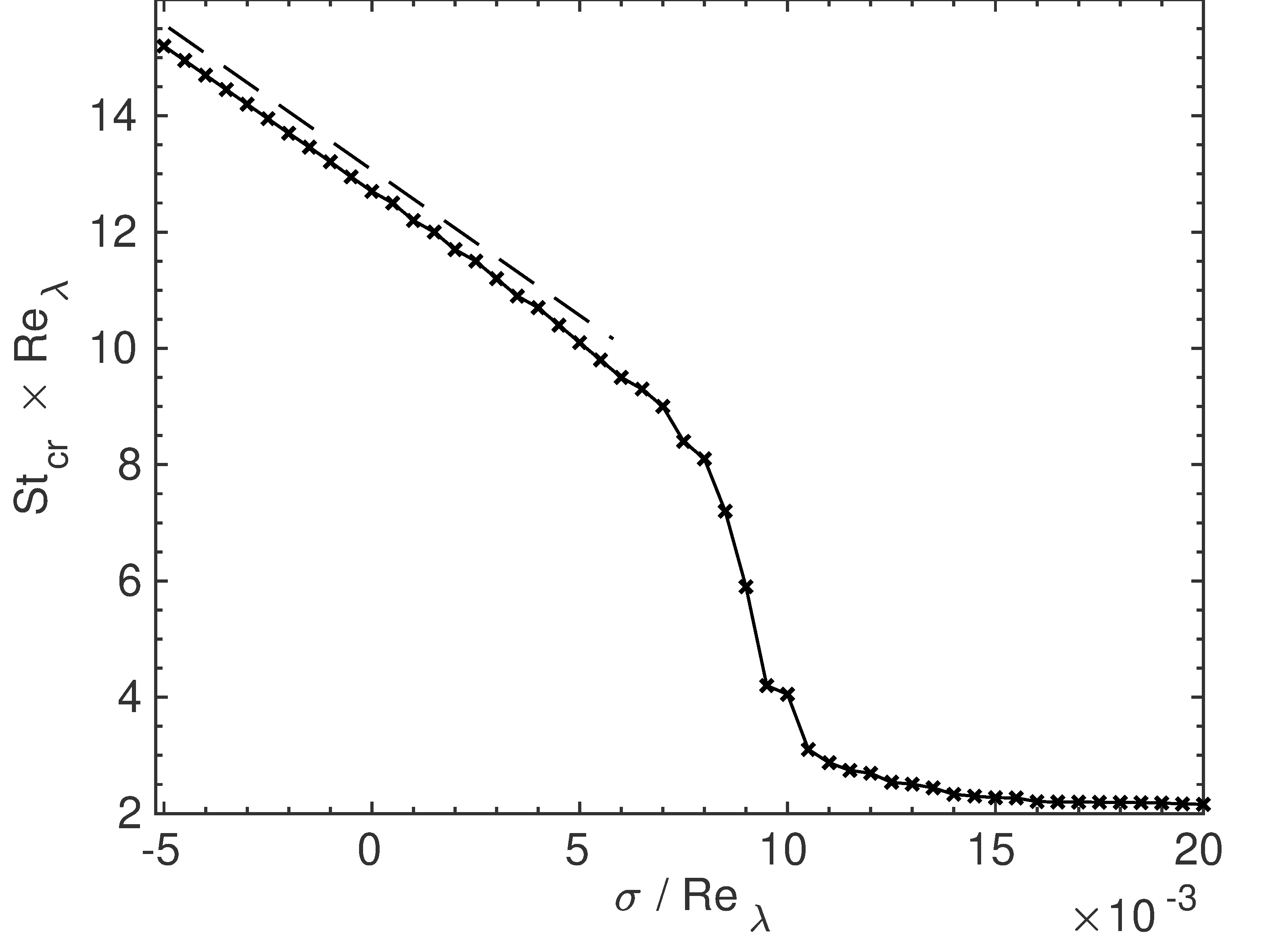}
\caption{The critical Stokes as a function of the nondimensional stretching rate. See, also, Fig.~\ref{fig:Stcr}}.
\label{fig:Stcr_scalefree}
\end{figure}

{Adapting this phenomenology to the Burgers vortex, we set $r_{\rm core} = l =
\SI{7.54e-3}{\meter}$ and $\Gamma = \SI{0.104}{\meter \squared \per \second}$
and the stretching coefficient $\sigma = \frac{2\nu}{r_{\rm core}^2} =
\SI{5.2e-1}{\per \second}$~\cite{Davidson}. These parameters, which build our
stretched Burgers vortex model, are of course dimensional. However, it is
important for our numerical simulations to deal with dimensionless parameters.
Using $\tau_\eta$ as a time scale and $25 \eta$  as a length scale, we obtain the non-dimensional parameters for our
simulations as $\Gamma = 11.313$, $r_{\rm core} = 0.4$, and $\sigma = 0.02$ (corresponding to a normalized $\tau_\eta=1$).
This $r_{\rm core}$ sets our parameters for the large core vortex.  Our other
parameters, listed in the main text of the paper, are variations of these
obtained by keeping $\Gamma$ fixed and applying the scaling $\sigma \propto
1/{r_{\rm core}^2}$.  In some cases however, e.g. in Figs.~\ref{fig:radial} to
~\ref{fig:Phi}, we vary $\sigma$ while keeping $r_{\rm core}$ fixed, in order
to isolate the influence of the stretching flow.}

The particles are characterised by their Stokes time:
\begin{equation}
    \tau_p =\frac{2a^2\rho_p}{9\nu\rho_f}
\end{equation}
where $a$ is the radius of the particle, $\rho_p$ is the density of the
particle (liquid water), and $\rho_f$ is the density of the carrier fluid (air). Typical droplet nuclei in a warm 
cloud grow up to a size $a \approx \SI{10}{\micro \meter}$ through condensation; further 
rapid growth up to sizes 8 times larger, which can trigger rain, is due to coalescences.
By taking typical values of the densities, $\rho_f = \SI{1.225}{\kilo \gram \per \meter \cubed}$ and
$\rho_p = \SI{9.97e2}{\kilo \gram \per \meter \cubed}$, we obtain 
$\tau_p = \SI{1.22e-3}{\second}$. This yields a Stokes number 
$St = \frac{\tau_p}{\tau_{\eta}} = \SI{3.17e-2}{}$. In our simulations, we work with Stokes numbers that match this typical value, by suitably selecting values of 
$\tau_p$, keeping in mind the normalization $\tau_\eta = 1.0$ (see above). The values of $St_{cr}$ presented in Fig.~\ref{fig:Stcr} fall within
this range of $St$, indicating that even such small particles can
undergo caustics in the vicinity of strongly stretched, intense
vortices. Fig.~\ref{fig:Stcr} was calculated for the case of $Re_\lambda = 5000$.
For geophysical flows with higher values of $Re_\lambda$, we expect
this effect to be further enhanced, as more intense vortices are
formed. This dependence on $Re_\lambda$ is made apparent in Fig.~\ref{fig:Stcr_scalefree},
which is a generalized version of Fig.~\ref{fig:Stcr}.


\bibliography{ref_turb_part}

\begin{thebibliography}{42}%
\makeatletter
\providecommand \@ifxundefined [1]{%
 \@ifx{#1\undefined}
}%
\providecommand \@ifnum [1]{%
 \ifnum #1\expandafter \@firstoftwo
 \else \expandafter \@secondoftwo
 \fi
}%
\providecommand \@ifx [1]{%
 \ifx #1\expandafter \@firstoftwo
 \else \expandafter \@secondoftwo
 \fi
}%
\providecommand \natexlab [1]{#1}%
\providecommand \enquote  [1]{``#1''}%
\providecommand \bibnamefont  [1]{#1}%
\providecommand \bibfnamefont [1]{#1}%
\providecommand \citenamefont [1]{#1}%
\providecommand \href@noop [0]{\@secondoftwo}%
\providecommand \href [0]{\begingroup \@sanitize@url \@href}%
\providecommand \@href[1]{\@@startlink{#1}\@@href}%
\providecommand \@@href[1]{\endgroup#1\@@endlink}%
\providecommand \@sanitize@url [0]{\catcode `\\12\catcode `\$12\catcode
  `\&12\catcode `\#12\catcode `\^12\catcode `\_12\catcode `\%12\relax}%
\providecommand \@@startlink[1]{}%
\providecommand \@@endlink[0]{}%
\providecommand \url  [0]{\begingroup\@sanitize@url \@url }%
\providecommand \@url [1]{\endgroup\@href {#1}{\urlprefix }}%
\providecommand \urlprefix  [0]{URL }%
\providecommand \Eprint [0]{\href }%
\providecommand \doibase [0]{http://dx.doi.org/}%
\providecommand \selectlanguage [0]{\@gobble}%
\providecommand \bibinfo  [0]{\@secondoftwo}%
\providecommand \bibfield  [0]{\@secondoftwo}%
\providecommand \translation [1]{[#1]}%
\providecommand \BibitemOpen [0]{}%
\providecommand \bibitemStop [0]{}%
\providecommand \bibitemNoStop [0]{.\EOS\space}%
\providecommand \EOS [0]{\spacefactor3000\relax}%
\providecommand \BibitemShut  [1]{\csname bibitem#1\endcsname}%
\let\auto@bib@innerbib\@empty
\bibitem [{\citenamefont {Bec}(2003)}]{Bec2003}%
  \BibitemOpen
  \bibfield  {author} {\bibinfo {author} {\bibfnamefont {J.}~\bibnamefont
  {Bec}},\ }\href {\doibase 10.1063/1.1612500} {\bibfield  {journal} {\bibinfo
  {journal} {Physics of Fluids}\ }\textbf {\bibinfo {volume} {15}},\ \bibinfo
  {pages} {1} (\bibinfo {year} {2003})}\BibitemShut {NoStop}%
\bibitem [{\citenamefont {Beard}\ and\ \citenamefont
  {Ochs}(1993)}]{doi:10.1175/1520-0450(1993)032<0608:WRIAOO>2.0.CO;2}%
  \BibitemOpen
  \bibfield  {author} {\bibinfo {author} {\bibfnamefont {K.~V.}\ \bibnamefont
  {Beard}}\ and\ \bibinfo {author} {\bibfnamefont {H.~T.}\ \bibnamefont
  {Ochs}},\ }\href {\doibase 10.1175/1520-0450(1993)032<0608:WRIAOO>2.0.CO;2}
  {\bibfield  {journal} {\bibinfo  {journal} {J. App. Met.}\ }\textbf {\bibinfo
  {volume} {32}},\ \bibinfo {pages} {608} (\bibinfo {year} {1993})}\BibitemShut
  {NoStop}%
\bibitem [{\citenamefont {Bec}\ \emph {et~al.}(2014)\citenamefont {Bec},
  \citenamefont {Homann},\ and\ \citenamefont {Ray}}]{PhysRevLett.112.184501}%
  \BibitemOpen
  \bibfield  {author} {\bibinfo {author} {\bibfnamefont {J.}~\bibnamefont
  {Bec}}, \bibinfo {author} {\bibfnamefont {H.}~\bibnamefont {Homann}}, \ and\
  \bibinfo {author} {\bibfnamefont {S.~S.}\ \bibnamefont {Ray}},\ }\href
  {\doibase 10.1103/PhysRevLett.112.184501} {\bibfield  {journal} {\bibinfo
  {journal} {Phys. Rev. Lett.}\ }\textbf {\bibinfo {volume} {112}},\ \bibinfo
  {pages} {184501} (\bibinfo {year} {2014})}\BibitemShut {NoStop}%
\bibitem [{\citenamefont {Bec}\ \emph {et~al.}(2016)\citenamefont {Bec},
  \citenamefont {Ray}, \citenamefont {Saw},\ and\ \citenamefont
  {Homann}}]{PhysRevE.93.031102}%
  \BibitemOpen
  \bibfield  {author} {\bibinfo {author} {\bibfnamefont {J.}~\bibnamefont
  {Bec}}, \bibinfo {author} {\bibfnamefont {S.~S.}\ \bibnamefont {Ray}},
  \bibinfo {author} {\bibfnamefont {E.~W.}\ \bibnamefont {Saw}}, \ and\
  \bibinfo {author} {\bibfnamefont {H.}~\bibnamefont {Homann}},\ }\href
  {\doibase 10.1103/PhysRevE.93.031102} {\bibfield  {journal} {\bibinfo
  {journal} {Phys. Rev. E}\ }\textbf {\bibinfo {volume} {93}},\ \bibinfo
  {pages} {031102} (\bibinfo {year} {2016})}\BibitemShut {NoStop}%
\bibitem [{\citenamefont {Borgas}\ and\ \citenamefont
  {Sawford}(1994)}]{borgas_sawford_1994}%
  \BibitemOpen
  \bibfield  {author} {\bibinfo {author} {\bibfnamefont {M.~S.}\ \bibnamefont
  {Borgas}}\ and\ \bibinfo {author} {\bibfnamefont {B.~L.}\ \bibnamefont
  {Sawford}},\ }\href {\doibase 10.1017/S0022112094003824} {\bibfield
  {journal} {\bibinfo  {journal} {J. Fluid Mech.}\ }\textbf {\bibinfo {volume}
  {279}},\ \bibinfo {pages} {69–99} (\bibinfo {year} {1994})}\BibitemShut
  {NoStop}%
\bibitem [{\citenamefont {Ravichandan}\ \emph {et~al.}(2017)\citenamefont
  {Ravichandan}, \citenamefont {Deepu},\ and\ \citenamefont
  {Govindarajan}}]{Rama-review}%
  \BibitemOpen
  \bibfield  {author} {\bibinfo {author} {\bibfnamefont {S.}~\bibnamefont
  {Ravichandan}}, \bibinfo {author} {\bibfnamefont {P.}~\bibnamefont {Deepu}},
  \ and\ \bibinfo {author} {\bibfnamefont {R.}~\bibnamefont {Govindarajan}},\
  }\href@noop {} {\bibfield  {journal} {\bibinfo  {journal} {Sadhana}\ }\textbf
  {\bibinfo {volume} {42}},\ \bibinfo {pages} {597} (\bibinfo {year}
  {2017})}\BibitemShut {NoStop}%
\bibitem [{\citenamefont {Johansen}\ \emph {et~al.}(2007)\citenamefont
  {Johansen}, \citenamefont {Oishi}, \citenamefont {Low}, \citenamefont
  {Klahr}, \citenamefont {Henning},\ and\ \citenamefont
  {Youdin}}]{Planets-Nature}%
  \BibitemOpen
  \bibfield  {author} {\bibinfo {author} {\bibfnamefont {A.}~\bibnamefont
  {Johansen}}, \bibinfo {author} {\bibfnamefont {J.~S.}\ \bibnamefont {Oishi}},
  \bibinfo {author} {\bibfnamefont {M.-M.~M.}\ \bibnamefont {Low}}, \bibinfo
  {author} {\bibfnamefont {H.}~\bibnamefont {Klahr}}, \bibinfo {author}
  {\bibfnamefont {T.}~\bibnamefont {Henning}}, \ and\ \bibinfo {author}
  {\bibfnamefont {A.}~\bibnamefont {Youdin}},\ }\href
  {http://dx.doi.org/10.1038/nature06086} {\bibfield  {journal} {\bibinfo
  {journal} {Nature}\ }\textbf {\bibinfo {volume} {448}},\ \bibinfo {pages}
  {1022} (\bibinfo {year} {2007})}\BibitemShut {NoStop}%
\bibitem [{\citenamefont {Squires}\ and\ \citenamefont
  {Eaton}(1991)}]{doi:10.1063/1.858045}%
  \BibitemOpen
  \bibfield  {author} {\bibinfo {author} {\bibfnamefont {K.~D.}\ \bibnamefont
  {Squires}}\ and\ \bibinfo {author} {\bibfnamefont {J.~K.}\ \bibnamefont
  {Eaton}},\ }\href {\doibase 10.1063/1.858045} {\bibfield  {journal} {\bibinfo
   {journal} {Phys. Fluids}\ }\textbf {\bibinfo {volume} {3}},\ \bibinfo
  {pages} {1169} (\bibinfo {year} {1991})}\BibitemShut {NoStop}%
\bibitem [{\citenamefont {Wang}\ and\ \citenamefont
  {Maxey}(1993)}]{wang_maxey_1993}%
  \BibitemOpen
  \bibfield  {author} {\bibinfo {author} {\bibfnamefont {L.-P.}\ \bibnamefont
  {Wang}}\ and\ \bibinfo {author} {\bibfnamefont {M.~R.}\ \bibnamefont
  {Maxey}},\ }\href {\doibase 10.1017/S0022112093002708} {\bibfield  {journal}
  {\bibinfo  {journal} {J. Fluid Mech.}\ }\textbf {\bibinfo {volume} {256}},\
  \bibinfo {pages} {27–68} (\bibinfo {year} {1993})}\BibitemShut {NoStop}%
\bibitem [{\citenamefont {Wood}\ \emph {et~al.}(2005)\citenamefont {Wood},
  \citenamefont {Hwang},\ and\ \citenamefont {Eaton}}]{WOOD20051220}%
  \BibitemOpen
  \bibfield  {author} {\bibinfo {author} {\bibfnamefont {A.}~\bibnamefont
  {Wood}}, \bibinfo {author} {\bibfnamefont {W.}~\bibnamefont {Hwang}}, \ and\
  \bibinfo {author} {\bibfnamefont {J.}~\bibnamefont {Eaton}},\ }\href
  {\doibase https://doi.org/10.1016/j.ijmultiphaseflow.2005.07.001} {\bibfield
  {journal} {\bibinfo  {journal} {Int. J. of Multiphase Flow}\ }\textbf
  {\bibinfo {volume} {31}},\ \bibinfo {pages} {1220 } (\bibinfo {year}
  {2005})}\BibitemShut {NoStop}%
\bibitem [{\citenamefont {Balachandar}\ and\ \citenamefont
  {Eaton}(2010)}]{doi:10.1146/annurev.fluid.010908.165243}%
  \BibitemOpen
  \bibfield  {author} {\bibinfo {author} {\bibfnamefont {S.}~\bibnamefont
  {Balachandar}}\ and\ \bibinfo {author} {\bibfnamefont {J.~K.}\ \bibnamefont
  {Eaton}},\ }\href@noop {} {\bibfield  {journal} {\bibinfo  {journal} {Annu.
  Rev. Fluid Mech.}\ }\textbf {\bibinfo {volume} {42}},\ \bibinfo {pages} {111}
  (\bibinfo {year} {2010})}\BibitemShut {NoStop}%
\bibitem [{\citenamefont
  {Shaw}(2003)}]{doi:10.1146/annurev.fluid.35.101101.161125}%
  \BibitemOpen
  \bibfield  {author} {\bibinfo {author} {\bibfnamefont {R.~A.}\ \bibnamefont
  {Shaw}},\ }\href@noop {} {\bibfield  {journal} {\bibinfo  {journal} {Annu.
  Rev. Fluid Mech.}\ }\textbf {\bibinfo {volume} {35}},\ \bibinfo {pages} {183}
  (\bibinfo {year} {2003})}\BibitemShut {NoStop}%
\bibitem [{\citenamefont {Bec}\ \emph {et~al.}(2005)\citenamefont {Bec},
  \citenamefont {Celani}, \citenamefont {Cencini},\ and\ \citenamefont
  {Musacchio}}]{Bec2005}%
  \BibitemOpen
  \bibfield  {author} {\bibinfo {author} {\bibfnamefont {J.}~\bibnamefont
  {Bec}}, \bibinfo {author} {\bibfnamefont {A.}~\bibnamefont {Celani}},
  \bibinfo {author} {\bibfnamefont {M.}~\bibnamefont {Cencini}}, \ and\
  \bibinfo {author} {\bibfnamefont {S.}~\bibnamefont {Musacchio}},\ }\href
  {\doibase 10.1063/1.1940367} {\bibfield  {journal} {\bibinfo  {journal}
  {Phys. Fluids}\ }\textbf {\bibinfo {volume} {17}},\ \bibinfo {pages} {1}
  (\bibinfo {year} {2005})}\BibitemShut {NoStop}%
\bibitem [{\citenamefont {Falkovich}\ and\ \citenamefont
  {Pumir}(2007)}]{Falkovich2007}%
  \BibitemOpen
  \bibfield  {author} {\bibinfo {author} {\bibfnamefont {G.}~\bibnamefont
  {Falkovich}}\ and\ \bibinfo {author} {\bibfnamefont {A.}~\bibnamefont
  {Pumir}},\ }\href@noop {} {\bibfield  {journal} {\bibinfo  {journal} {J.
  Atmos. Sciences}\ }\textbf {\bibinfo {volume} {64}},\ \bibinfo {pages} {4497}
  (\bibinfo {year} {2007})}\BibitemShut {NoStop}%
\bibitem [{\citenamefont {Bec}\ \emph {et~al.}(2013)\citenamefont {Bec},
  \citenamefont {Musacchio},\ and\ \citenamefont {Ray}}]{PhysRevE.87.063013}%
  \BibitemOpen
  \bibfield  {author} {\bibinfo {author} {\bibfnamefont {J.}~\bibnamefont
  {Bec}}, \bibinfo {author} {\bibfnamefont {S.}~\bibnamefont {Musacchio}}, \
  and\ \bibinfo {author} {\bibfnamefont {S.~S.}\ \bibnamefont {Ray}},\ }\href
  {\doibase 10.1103/PhysRevE.87.063013} {\bibfield  {journal} {\bibinfo
  {journal} {Phys. Rev. E}\ }\textbf {\bibinfo {volume} {87}},\ \bibinfo
  {pages} {063013} (\bibinfo {year} {2013})}\BibitemShut {NoStop}%
\bibitem [{\citenamefont {Pumir}\ and\ \citenamefont
  {Wilkinson}(2016)}]{Pumir2016}%
  \BibitemOpen
  \bibfield  {author} {\bibinfo {author} {\bibfnamefont {A.}~\bibnamefont
  {Pumir}}\ and\ \bibinfo {author} {\bibfnamefont {M.}~\bibnamefont
  {Wilkinson}},\ }\href {\doibase 10.1146/annurev-conmatphys-031115-011538}
  {\bibfield  {journal} {\bibinfo  {journal} {Annu. Rev. Cond. Matt. Phys.}\
  }\textbf {\bibinfo {volume} {7}},\ \bibinfo {pages} {141} (\bibinfo {year}
  {2016})},\ \Eprint {http://arxiv.org/abs/1508.01538} {1508.01538}
  \BibitemShut {NoStop}%
\bibitem [{\citenamefont {James}\ and\ \citenamefont {Ray}(2017)}]{JamesRay}%
  \BibitemOpen
  \bibfield  {author} {\bibinfo {author} {\bibfnamefont {M.}~\bibnamefont
  {James}}\ and\ \bibinfo {author} {\bibfnamefont {S.~S.}\ \bibnamefont
  {Ray}},\ }\href {\doibase 10.1038/s41598-017-12093-0} {\bibfield  {journal}
  {\bibinfo  {journal} {Sci. Rep.}\ }\textbf {\bibinfo {volume} {7}},\ \bibinfo
  {pages} {12231} (\bibinfo {year} {2017})}\BibitemShut {NoStop}%
\bibitem [{\citenamefont {Saw}\ \emph {et~al.}(2014)\citenamefont {Saw},
  \citenamefont {Bewley}, \citenamefont {Bodenschatz}, \citenamefont {Ray},\
  and\ \citenamefont {Bec}}]{doi:10.1063/1.4900848}%
  \BibitemOpen
  \bibfield  {author} {\bibinfo {author} {\bibfnamefont {E.-W.}\ \bibnamefont
  {Saw}}, \bibinfo {author} {\bibfnamefont {G.~P.}\ \bibnamefont {Bewley}},
  \bibinfo {author} {\bibfnamefont {E.}~\bibnamefont {Bodenschatz}}, \bibinfo
  {author} {\bibfnamefont {S.~S.}\ \bibnamefont {Ray}}, \ and\ \bibinfo
  {author} {\bibfnamefont {J.}~\bibnamefont {Bec}},\ }\href {\doibase
  10.1063/1.4900848} {\bibfield  {journal} {\bibinfo  {journal} {Phys. Fluids}\
  }\textbf {\bibinfo {volume} {26}},\ \bibinfo {pages} {111702} (\bibinfo
  {year} {2014})}\BibitemShut {NoStop}%
\bibitem [{\citenamefont {Picardo}\ \emph {et~al.}(2019)\citenamefont
  {Picardo}, \citenamefont {Agasthya}, \citenamefont {Govindarajan},\ and\
  \citenamefont {Ray}}]{Picardo2018-coll}%
  \BibitemOpen
  \bibfield  {author} {\bibinfo {author} {\bibfnamefont {J.~R.}\ \bibnamefont
  {Picardo}}, \bibinfo {author} {\bibfnamefont {L.}~\bibnamefont {Agasthya}},
  \bibinfo {author} {\bibfnamefont {R.}~\bibnamefont {Govindarajan}}, \ and\
  \bibinfo {author} {\bibfnamefont {S.~S.}\ \bibnamefont {Ray}},\ }\href
  {\doibase 10.1103/PhysRevFluids.4.032601} {\bibfield  {journal} {\bibinfo
  {journal} {Phys. Rev. Fluids}\ }\textbf {\bibinfo {volume} {4}},\ \bibinfo
  {pages} {032601 (R)} (\bibinfo {year} {2019})}\BibitemShut {NoStop}%
\bibitem [{\citenamefont {Deepu}\ \emph {et~al.}(2017)\citenamefont {Deepu},
  \citenamefont {Ravichandran},\ and\ \citenamefont
  {Govindarajan}}]{Deepu2017}%
  \BibitemOpen
  \bibfield  {author} {\bibinfo {author} {\bibfnamefont {P.}~\bibnamefont
  {Deepu}}, \bibinfo {author} {\bibfnamefont {S.}~\bibnamefont {Ravichandran}},
  \ and\ \bibinfo {author} {\bibfnamefont {R.}~\bibnamefont {Govindarajan}},\
  }\href@noop {} {\bibfield  {journal} {\bibinfo  {journal} {Phys. Rev.
  Fluids}\ }\textbf {\bibinfo {volume} {2}},\ \bibinfo {pages} {024305}
  (\bibinfo {year} {2017})}\BibitemShut {NoStop}%
\bibitem [{\citenamefont {Maxey}(1987)}]{maxey_1987}%
  \BibitemOpen
  \bibfield  {author} {\bibinfo {author} {\bibfnamefont {M.~R.}\ \bibnamefont
  {Maxey}},\ }\href {\doibase 10.1017/S0022112087000193} {\bibfield  {journal}
  {\bibinfo  {journal} {J. Fluid Mech.}\ }\textbf {\bibinfo {volume} {174}},\
  \bibinfo {pages} {441–465} (\bibinfo {year} {1987})}\BibitemShut {NoStop}%
\bibitem [{\citenamefont {Falkovich}\ \emph {et~al.}(2002)\citenamefont
  {Falkovich}, \citenamefont {Fouxon},\ and\ \citenamefont
  {Stepanov}}]{Falkovich2002}%
  \BibitemOpen
  \bibfield  {author} {\bibinfo {author} {\bibfnamefont {G.}~\bibnamefont
  {Falkovich}}, \bibinfo {author} {\bibfnamefont {A.}~\bibnamefont {Fouxon}}, \
  and\ \bibinfo {author} {\bibfnamefont {M.~G.}\ \bibnamefont {Stepanov}},\
  }\href@noop {} {\bibfield  {journal} {\bibinfo  {journal} {Nature}\ }\textbf
  {\bibinfo {volume} {419}},\ \bibinfo {pages} {151} (\bibinfo {year}
  {2002})}\BibitemShut {NoStop}%
\bibitem [{\citenamefont {Wilkinson}\ \emph {et~al.}(2006)\citenamefont
  {Wilkinson}, \citenamefont {Mehlig},\ and\ \citenamefont
  {Bezuglyy}}]{Wilkinson2006}%
  \BibitemOpen
  \bibfield  {author} {\bibinfo {author} {\bibfnamefont {M.}~\bibnamefont
  {Wilkinson}}, \bibinfo {author} {\bibfnamefont {B.}~\bibnamefont {Mehlig}}, \
  and\ \bibinfo {author} {\bibfnamefont {V.}~\bibnamefont {Bezuglyy}},\
  }\href@noop {} {\bibfield  {journal} {\bibinfo  {journal} {Phys. Rev. Lett.}\
  }\textbf {\bibinfo {volume} {97}},\ \bibinfo {pages} {048501} (\bibinfo
  {year} {2006})}\BibitemShut {NoStop}%
\bibitem [{\citenamefont {Bewley}\ \emph {et~al.}(2013)\citenamefont {Bewley},
  \citenamefont {Saw},\ and\ \citenamefont {Bodenschatz}}]{Bewley2013}%
  \BibitemOpen
  \bibfield  {author} {\bibinfo {author} {\bibfnamefont {G.~P.}\ \bibnamefont
  {Bewley}}, \bibinfo {author} {\bibfnamefont {E.-W.}\ \bibnamefont {Saw}}, \
  and\ \bibinfo {author} {\bibfnamefont {E.}~\bibnamefont {Bodenschatz}},\
  }\href@noop {} {\bibfield  {journal} {\bibinfo  {journal} {New J. Phys.}\
  }\textbf {\bibinfo {volume} {15}},\ \bibinfo {pages} {083051} (\bibinfo
  {year} {2013})}\BibitemShut {NoStop}%
\bibitem [{\citenamefont {Ravichandran}\ and\ \citenamefont
  {Govindarajan}(2015)}]{Ravichandran2015}%
  \BibitemOpen
  \bibfield  {author} {\bibinfo {author} {\bibfnamefont {S.}~\bibnamefont
  {Ravichandran}}\ and\ \bibinfo {author} {\bibfnamefont {R.}~\bibnamefont
  {Govindarajan}},\ }\href@noop {} {\bibfield  {journal} {\bibinfo  {journal}
  {Phys. Fluids}\ }\textbf {\bibinfo {volume} {27}} (\bibinfo {year}
  {2015})}\BibitemShut {NoStop}%
\bibitem [{\citenamefont {Burgers}(1948)}]{BURGERS1948171}%
  \BibitemOpen
  \bibfield  {author} {\bibinfo {author} {\bibfnamefont {J.~M.}\ \bibnamefont
  {Burgers}},\ }\href@noop {} {\emph {\bibinfo {title} {Advances in Applied
  Mechanics}}}\ (\bibinfo  {publisher} {Academic, New York, 1948},\ \bibinfo
  {year} {1948})\BibitemShut {NoStop}%
\bibitem [{\citenamefont {She}\ \emph {et~al.}(1990)\citenamefont {She},
  \citenamefont {Jackson},\ and\ \citenamefont {Orszag}}]{she1990}%
  \BibitemOpen
  \bibfield  {author} {\bibinfo {author} {\bibfnamefont {Z.}~\bibnamefont
  {She}}, \bibinfo {author} {\bibfnamefont {E.}~\bibnamefont {Jackson}}, \ and\
  \bibinfo {author} {\bibfnamefont {S.~A.}\ \bibnamefont {Orszag}},\
  }\href@noop {} {\bibfield  {journal} {\bibinfo  {journal} {Nature}\ }\textbf
  {\bibinfo {volume} {344}},\ \bibinfo {pages} {226} (\bibinfo {year}
  {1990})}\BibitemShut {NoStop}%
\bibitem [{\citenamefont {Jim{\'{e}}nez}\ and\ \citenamefont
  {Wray}(1998)}]{Jimenez1998}%
  \BibitemOpen
  \bibfield  {author} {\bibinfo {author} {\bibfnamefont {J.}~\bibnamefont
  {Jim{\'{e}}nez}}\ and\ \bibinfo {author} {\bibfnamefont {A.~A.}\ \bibnamefont
  {Wray}},\ }\href {\doibase 10.1017/S0022112098002341} {\bibfield  {journal}
  {\bibinfo  {journal} {J. Fluid Mech.}\ }\textbf {\bibinfo {volume} {373}},\
  \bibinfo {pages} {255} (\bibinfo {year} {1998})}\BibitemShut {NoStop}%
\bibitem [{\citenamefont {Ishihara}\ \emph {et~al.}(2007)\citenamefont
  {Ishihara}, \citenamefont {Kaneda}, \citenamefont {Yokokawa}, \citenamefont
  {Itakura},\ and\ \citenamefont {Uno}}]{ishihara2007}%
  \BibitemOpen
  \bibfield  {author} {\bibinfo {author} {\bibfnamefont {T.}~\bibnamefont
  {Ishihara}}, \bibinfo {author} {\bibfnamefont {Y.}~\bibnamefont {Kaneda}},
  \bibinfo {author} {\bibfnamefont {M.}~\bibnamefont {Yokokawa}}, \bibinfo
  {author} {\bibfnamefont {K.}~\bibnamefont {Itakura}}, \ and\ \bibinfo
  {author} {\bibfnamefont {A.}~\bibnamefont {Uno}},\ }\href {\doibase
  10.1017/S0022112007008531} {\bibfield  {journal} {\bibinfo  {journal} {J.
  Fluid Mech.}\ }\textbf {\bibinfo {volume} {592}},\ \bibinfo {pages}
  {335–366} (\bibinfo {year} {2007})}\BibitemShut {NoStop}%
\bibitem [{\citenamefont {Douady}\ \emph {et~al.}(1991)\citenamefont {Douady},
  \citenamefont {Couder},\ and\ \citenamefont {Brachet}}]{Douady1991}%
  \BibitemOpen
  \bibfield  {author} {\bibinfo {author} {\bibfnamefont {S.}~\bibnamefont
  {Douady}}, \bibinfo {author} {\bibfnamefont {Y.}~\bibnamefont {Couder}}, \
  and\ \bibinfo {author} {\bibfnamefont {M.~E.}\ \bibnamefont {Brachet}},\
  }\href {\doibase 10.1103/PhysRevLett.67.983} {\bibfield  {journal} {\bibinfo
  {journal} {Phys. Rev. Lett.}\ }\textbf {\bibinfo {volume} {67}},\ \bibinfo
  {pages} {983} (\bibinfo {year} {1991})}\BibitemShut {NoStop}%
\bibitem [{\citenamefont {Lundgren}(1982)}]{doi:10.1063/1.863957}%
  \BibitemOpen
  \bibfield  {author} {\bibinfo {author} {\bibfnamefont {T.~S.}\ \bibnamefont
  {Lundgren}},\ }\href {\doibase 10.1063/1.863957} {\bibfield  {journal}
  {\bibinfo  {journal} {Phys. Fluids}\ }\textbf {\bibinfo {volume} {25}},\
  \bibinfo {pages} {2193} (\bibinfo {year} {1982})}\BibitemShut {NoStop}%
\bibitem [{\citenamefont {Gibbon}\ \emph {et~al.}(1999)\citenamefont {Gibbon},
  \citenamefont {Fokas},\ and\ \citenamefont {Doering}}]{GIBBON1999497}%
  \BibitemOpen
  \bibfield  {author} {\bibinfo {author} {\bibfnamefont {J.}~\bibnamefont
  {Gibbon}}, \bibinfo {author} {\bibfnamefont {A.}~\bibnamefont {Fokas}}, \
  and\ \bibinfo {author} {\bibfnamefont {C.}~\bibnamefont {Doering}},\
  }\href@noop {} {\bibfield  {journal} {\bibinfo  {journal} {Physica D}\
  }\textbf {\bibinfo {volume} {132}},\ \bibinfo {pages} {497 } (\bibinfo {year}
  {1999})}\BibitemShut {NoStop}%
\bibitem [{\citenamefont {Galanti}\ \emph {et~al.}(1997)\citenamefont
  {Galanti}, \citenamefont {Gibbon},\ and\ \citenamefont
  {Heritage}}]{Galanti:1997}%
  \BibitemOpen
  \bibfield  {author} {\bibinfo {author} {\bibfnamefont {B.}~\bibnamefont
  {Galanti}}, \bibinfo {author} {\bibfnamefont {J.}~\bibnamefont {Gibbon}}, \
  and\ \bibinfo {author} {\bibfnamefont {M.}~\bibnamefont {Heritage}},\
  }\href@noop {} {\bibfield  {journal} {\bibinfo  {journal} {Nonlinearity}\
  }\textbf {\bibinfo {volume} {10}},\ \bibinfo {pages} {1675} (\bibinfo {year}
  {1997})}\BibitemShut {NoStop}%
\bibitem [{\citenamefont {Gibbon}\ and\ \citenamefont
  {Heritage}(1997)}]{Gibbon:1997:10.1063/1.869186}%
  \BibitemOpen
  \bibfield  {author} {\bibinfo {author} {\bibfnamefont {J.}~\bibnamefont
  {Gibbon}}\ and\ \bibinfo {author} {\bibfnamefont {M.}~\bibnamefont
  {Heritage}},\ }\href {\doibase 10.1063/1.869186} {\bibfield  {journal}
  {\bibinfo  {journal} {Phys. Fluids}\ }\textbf {\bibinfo {volume} {9}},\
  \bibinfo {pages} {901} (\bibinfo {year} {1997})}\BibitemShut {NoStop}%
\bibitem [{\citenamefont {Chapman}\ \emph {et~al.}(1997)\citenamefont
  {Chapman}, \citenamefont {Elliott}, \citenamefont {Head}, \citenamefont
  {Howison}, \citenamefont {Leslie}, \citenamefont {Ockendon},\ and\
  \citenamefont {Saffman}}]{Chapman}%
  \BibitemOpen
  \bibfield  {author} {\bibinfo {author} {\bibfnamefont {S.~J.}\ \bibnamefont
  {Chapman}}, \bibinfo {author} {\bibfnamefont {C.~M.}\ \bibnamefont
  {Elliott}}, \bibinfo {author} {\bibfnamefont {A.~K.}\ \bibnamefont {Head}},
  \bibinfo {author} {\bibfnamefont {S.~D.}\ \bibnamefont {Howison}}, \bibinfo
  {author} {\bibfnamefont {F.~M.}\ \bibnamefont {Leslie}}, \bibinfo {author}
  {\bibfnamefont {J.~R.}\ \bibnamefont {Ockendon}}, \ and\ \bibinfo {author}
  {\bibfnamefont {P.~G.}\ \bibnamefont {Saffman}},\ }\href {\doibase
  10.1098/rsta.1997.0097} {\bibfield  {journal} {\bibinfo  {journal}
  {Philosophical Transactions of the Royal Society of London. Series A:
  Mathematical, Physical and Engineering Sciences}\ }\textbf {\bibinfo {volume}
  {355}},\ \bibinfo {pages} {1949} (\bibinfo {year} {1997})}\BibitemShut
  {NoStop}%
\bibitem [{\citenamefont {Wilczek}\ \emph {et~al.}(2008)\citenamefont
  {Wilczek}, \citenamefont {Jenko},\ and\ \citenamefont {Friedrich}}]{Wilczek}%
  \BibitemOpen
  \bibfield  {author} {\bibinfo {author} {\bibfnamefont {M.}~\bibnamefont
  {Wilczek}}, \bibinfo {author} {\bibfnamefont {F.}~\bibnamefont {Jenko}}, \
  and\ \bibinfo {author} {\bibfnamefont {R.}~\bibnamefont {Friedrich}},\
  }\href@noop {} {\bibfield  {journal} {\bibinfo  {journal} {Phys. Rev. E}\
  }\textbf {\bibinfo {volume} {77}},\ \bibinfo {pages} {056301} (\bibinfo
  {year} {2008})}\BibitemShut {NoStop}%
\bibitem [{\citenamefont {Marcu}\ \emph {et~al.}(1995)\citenamefont {Marcu},
  \citenamefont {Meiburg},\ and\ \citenamefont {Newton}}]{Marcu1998}%
  \BibitemOpen
  \bibfield  {author} {\bibinfo {author} {\bibfnamefont {B.}~\bibnamefont
  {Marcu}}, \bibinfo {author} {\bibfnamefont {E.}~\bibnamefont {Meiburg}}, \
  and\ \bibinfo {author} {\bibfnamefont {P.~K.}\ \bibnamefont {Newton}},\
  }\href@noop {} {\bibfield  {journal} {\bibinfo  {journal} {Physics of
  Fluids}\ }\textbf {\bibinfo {volume} {7}},\ \bibinfo {pages} {400} (\bibinfo
  {year} {1995})}\BibitemShut {NoStop}%
\bibitem [{\citenamefont {Hill}(2005)}]{Hill}%
  \BibitemOpen
  \bibfield  {author} {\bibinfo {author} {\bibfnamefont {R.~J.}\ \bibnamefont
  {Hill}},\ }\href@noop {} {\bibfield  {journal} {\bibinfo  {journal} {Phys.
  Fluids}\ }\textbf {\bibinfo {volume} {17}},\ \bibinfo {pages} {037103}
  (\bibinfo {year} {2005})}\BibitemShut {NoStop}%
\bibitem [{\citenamefont {Davidson}(2004)}]{Davidson}%
  \BibitemOpen
  \bibfield  {author} {\bibinfo {author} {\bibfnamefont {P.~A.}\ \bibnamefont
  {Davidson}},\ }\href@noop {} {\emph {\bibinfo {title} {Turbulence: An
  Introduction for Scientists and Engineers}}}\ (\bibinfo  {publisher} {Oxford
  University Press, UK},\ \bibinfo {year} {2004})\BibitemShut {NoStop}%
\bibitem [{\citenamefont {Maxey}(1983)}]{Maxey1983}%
  \BibitemOpen
  \bibfield  {author} {\bibinfo {author} {\bibfnamefont {M.~R.}\ \bibnamefont
  {Maxey}},\ }\href@noop {} {\bibfield  {journal} {\bibinfo  {journal} {Phys.
  Fluids}\ }\textbf {\bibinfo {volume} {26}},\ \bibinfo {pages} {883} (\bibinfo
  {year} {1983})}\BibitemShut {NoStop}%
\bibitem [{\citenamefont {Sundaram}\ and\ \citenamefont
  {Collins}(1997)}]{sundaram_collins_1997}%
  \BibitemOpen
  \bibfield  {author} {\bibinfo {author} {\bibfnamefont {S.}~\bibnamefont
  {Sundaram}}\ and\ \bibinfo {author} {\bibfnamefont {L.~R.}\ \bibnamefont
  {Collins}},\ }\href {\doibase 10.1017/S0022112097001833} {\bibfield
  {journal} {\bibinfo  {journal} {J. Fluid Mech.}\ }\textbf {\bibinfo {volume}
  {335}},\ \bibinfo {pages} {75–109} (\bibinfo {year} {1997})}\BibitemShut
  {NoStop}%
\bibitem [{\citenamefont {Bhatnagar}\ \emph {et~al.}(2018)\citenamefont
  {Bhatnagar}, \citenamefont {Gustavsson}, \citenamefont {Mehlig},\ and\
  \citenamefont {Mitra}}]{Bhatnagar}%
  \BibitemOpen
  \bibfield  {author} {\bibinfo {author} {\bibfnamefont {A.}~\bibnamefont
  {Bhatnagar}}, \bibinfo {author} {\bibfnamefont {K.}~\bibnamefont
  {Gustavsson}}, \bibinfo {author} {\bibfnamefont {B.}~\bibnamefont {Mehlig}},
  \ and\ \bibinfo {author} {\bibfnamefont {D.}~\bibnamefont {Mitra}},\ }\href
  {\doibase 10.1103/PhysRevE.98.063107} {\bibfield  {journal} {\bibinfo
  {journal} {Phys. Rev. E}\ }\textbf {\bibinfo {volume} {98}},\ \bibinfo
  {pages} {063107} (\bibinfo {year} {2018})}\BibitemShut {NoStop}%
\end{thebibliography}%

\end{document}